# The Resiliency of Memorability: A Predictor of Memory Separate from Attention and Priming

Wilma A. Bainbridge

Department of Brain and Cognitive Sciences, Massachusetts Institute of Technology. Cambridge, MA. USA.

**ABSTRACT:** When we encounter a new person or place, we may easily encode it into our memories, or we may quickly forget it. Recent work finds that this likelihood of encoding a given entity – memorability – is highly consistent across viewers and intrinsic to an image; people tend to remember and forget the same images. However, several forces influence our memories beyond the memorability of the stimulus itself – for example, how attention-grabbing the stimulus is, how much attentional resources we dedicate to the task, or how primed we are for that stimulus. How does memorability interact with these various phenomena, and could any of them explain the effects of memorability found in prior work? This study uses five psychophysical experiments to explore the link between memorability and three attention-related phenomena: 1) bottom-up attention (through testing spatial cueing and visual search), 2) top-down attention (through testing cognitive control and depth of encoding), and 3) priming. These experiments find that memorability remains resilient to all of these phenomena – none are able to explain memorability effects or overcome the strong effects memorability has on determining memory performance. Thus, memorability is truly an independent, intrinsic attribute of an image that works in conjunction with these phenomena to determine if an event will ultimately be remembered.

One great mystery of the human experience is why our memories often act against our will – we remember events that are not particularly important to us, yet we forget the names and faces of new acquaintances that we try desperately to remember. Recent work has pinpointed a novel image attribute that can help explain what we ultimately remember – *memorability*, a predictive value of the likelihood of a novel event being eventually remembered or forgotten (Isola et al., 2011a). Despite our diverse unique experiences, we tend to remember the same scenes (Isola et al., 2011b), faces (Bainbridge et al., 2013; Bainbridge, 2016), and even visualizations (Borkin et al., 2013) as each other. A stimulus's memorability plays a main role in influencing our future memories – intrinsic stimulus memorability makes up 50% of the variance in memory performance (with observer characteristics, environment features, noise, etc, making up the other 50%; Bainbridge et al., 2013). Image memorability shows stereotyped perception- and memory-based activity in the brain (Bainbridge et al., in press) and can also be predicted by convolutional neural networks (Khosla et al., 2015). Additionally, memorability cannot be fully explained by a wide range of face and scene attributes (Isola et al., 2011a; Bainbridge et al., 2013), is surprisingly resilient to different time scales (Isola et al., 2013) contexts (Bylinskii et al., 2015), and transformations of the stimulus entity (Bainbridge, in 2016). With the combination of these results, memorability appears to be its own independent phenomenon and image attribute that can guide predictions of later memory.

However, one alternative hypothesis is that these memorability effects are instead driven by attention. This could be in the form of bottom-up attention (also called exogenous attention) or also top-down attention (endogenous attention). Attention and memory are well-known to have a symbiotic relationship; with attention influencing what is encoded into memory, and previous experience also influencing what is attended to (Chun & Turk-Browne, 2007). Certain types of stimuli are known to draw more attention than others (Asmundson & Stein, 1994; Theeuwes & Van der Stigchel, 2006; Cooper & Langton, 2006; Bar-Haim et al., 2007; Trawalter et al., 2008), and stimuli that are attended to are more easily remembered (MacLeod, 1989; Chun & Turk-Browne, 2007). Memorability is similarly consistent across stimuli and predictive of later memory behavior. Thus perhaps memorability is synonymous with bottom-up attention orienting, or perhaps some of the robust effects of memorability can be largely explained by top-down attentional effects. A final candidate explanation for memorability is priming, where perhaps the non-attentive effects on memory from memorability could be instead explained by degree of priming caused by a stimulus.

The current study explores the relationship between memorability and three attention-related phenomena: bottom-up attention, top-down attention, and non-attentive priming. Experiments 1 and 2 find memorability does not influence bottom-up attention orienting in a spatial cueing task (Experiment 1) nor in a visual search task (Experiment 2). Experiments 3 and 4 find memorability influences memory behavior separately from top-down attention effects such as cognitive control (Experiment 3) and deeper encoding (Experiment 4). Lastly, Experiment 5 finds that memorability also shows a separate effect from perceptual priming. Taken together, these results form powerful evidence that memorability is an isolated stimulus property that cannot be explained by bottom-up attention, top-down attention, or priming.

## Section 1: Memorability and Bottom-Up Attention

### Section Introduction

Just in the same way emotional or threatening stimuli cause automatic bottom-up attentional orienting (Asmundson & Stein, 1994; Theeuwes & Van der Stigchel, 2006; Cooper & Langton, 2006; Bar-Haim et al., 2007), it seems likely that highly memorable stimuli are memorable because they are attention-grabbing. Indeed, stimulus memorability has been found to be correlated with many of the same features that also cause bottom-up attentional orienting – emotion, threat, and contrasting colors and brightness (Isola et al., 2011; Bainbridge et al., 2013). Various works have also identified specific effects of *distinctive* stimuli on visual search tasks (Treisman & Gormican, 1988; Wolfe, 2001), and novel stimuli have been found to capture and bind attention in comparison to familiar stimuli (Horstmann & Herwig, 2015). While distinctiveness and novelty are not identical with memorability (Bainbridge et al., 2013), they are highly correlated. Together, it thus seems likely that memorability occurs because of visual features in that memorable stimulus that cause it to quickly and automatically capture attention.

This hypothesis is tested with two paradigms often used in exogenous attentional experiments – spatial cueing and visual search. In the two following experiments, I find that while I am able to replicate the classical results of each task, memorability is found to have no effect on bottom-up attention effects, providing evidence for memorability as an image property that does not necessarily capture bottom-up attention.

## Experiment 1: Memorability and Spatial Cueing

**Introduction**

One common exogenous attention paradigm that can be adapted for complex stimuli (i.e., photographs rather than shapes or colors) is the dot-probe paradigm, or spatial cueing task (Posner, 1980). Participants see two irrelevant visual cues on the left and right sides of a fixation cross, and then must categorize a target image that appears on either side. If one of the irrelevant cues captures attention, then participants will automatically orient and respond faster if the target also occurs on the same side (Posner, 1980). In the realm of faces, this paradigm has found attention orienting towards faces differing in emotion (Theeuwes & Van der Stigchel, 2006; Cooper & Langton, 2006), level of threat (Asmundson & Stein, 1994; Bar-Haim et al., 2007), and race (Trawalter et al., 2008). This task is thus well-suited to examine whether memorable faces similarly cause automatic spatial cueing. If memorable images capture attention, then seeing a memorable face before a target should speed up the response to the target.

Two spatial cueing experiments were conducted to examine the bottom-up attentional effects of memorability. The first experiment (Experiment 1-A) replicated previous spatial cueing work, using face images as cues, with half at a normal brightness level and the other half at a high brightness level (Johannes et al., 1995; Lupiáñez et al., 2004). The second experiment (Experiment 1-B) was a spatial cueing study using memorable and forgettable faces of equal brightness as the cues. While previous spatial cueing work was successfully replicated, memorability was not found to cause any spatial cueing effects.

**Materials and Methods**

*Participants*

Participants for the two experiments were recruited from online crowdsourcing platform Amazon Mechanical Turk (AMT). For this and all Experiments in the current study, data were collected following the standards of the MIT Institutional Review Board, and all participants provided consent for the study. There were 96 participants for Experiment 1-A and 98 participants for Experiment 1-B. This large number of participants was recruited to ensure any inability to detect an effect was not due to a small sample size, and unequal numbers of participants (for this Experiment and all following) were due to a small proportion of AMT workers accepting the experiment but quitting partway through (which is not uncommon in crowdsourced experiments: Eickhoff & de Vries, 2013). Two similar follow-up experiments were conducted with Experiment 1-B to confirm its effects, with 26 and 23 participants. All participants were compensated for their time. Only participants with over a 95% AMT approval rating and an IP address in the United States were recruited for the study, so that their exposure to different facial demographics would most closely match those of the stimulus set (designed to approximate the U.S. population).

*Stimuli*

Both experiments used the same base set of 80 face images, 40 determined to be highly memorable (top 25% of hit rate, HR; M=72.5%, SD=6.7%) and 40 determined to be highly forgettable (bottom 25% of HR; M=32.4%, SD=5.5%) in a previous large-scale online memory test (Bainbridge et al., 2013). The conditions were selected to have no difference in false alarm rate (FAR) and several low- and mid-level attributes (color, spatial frequency, emotion, attractiveness). Faces were also matched in memorable and forgettable pairs for gender, race, and age. All face images are 256 pixels in height and cropped with an oval to diminish background effects. For the brightened stimuli in Experiment 1-A, RGB values of the images were all uniformly increased by 140 points.

*Experimental Methods*

Both experiments were conducted with a spatial cueing paradigm designed based on previous spatial cueing work (See Fig. 1; Posner, 1980; MacLeod et al., 1986; Fox et al., 2008) implemented on online psychology experimental platform PsyToolkit (Stoet, 2010).

A fixation cross was presented for 300ms, and then cue face images were presented 150 pixels to the left and 150 pixels to the right of the fixation cross for 500ms. The face images disappeared and after a randomized delay ranging between 100ms and 500ms, two dots (the dot probe) were presented on one side of the fixation cross for up to 1500ms. The randomized delay was used so that participants had to remain attentive during the whole trial, and could not anticipate when the target would appear while ignoring the cues. Participants were asked to respond as quickly and accurately as possible whether the dots were oriented horizontally or vertically, using a keyboard press that captured reaction time (RT). The position and orientation of the dot probe were randomly counterbalanced across trials. The face image cues had no relation to the dot probe. As soon as participants responded, the screen was cleared for 500ms and then continued onto the next trial. Participants were informed if their response was incorrect, and given their total correct at the end of the experiment. Each participant did 40 trials, of randomized order, and the experiment lasted approximately 3 minutes in total.

For Experiment 1-A, each trial had a brightened (low-contrast) face and a normal brightness face image cue, with memorability controlled for between conditions. It is expected that normal brightness cue should capture attention and result in faster RTs. For Experiment 1-B, each trial had a forgettable face image cue and a memorable face image cue, matched for demographics and of equal brightness. If memorability captures attention in a similar bottom-up way, then dot probes matching the side of the memorable image should have faster RTs. Analyses only used RTs from trials where participants gave the correct response on the trial, however no statistical tests change in significance if incorrect trial RTs are also included.

To ensure effects found from Experiment 1-B were not due to differences in trial timing, as there is debate in the literature in terms of the most effective spatial cueing task timing (Cooper & Langton, 2006), two almost identical follow-up studies to Experiment 1-B were also conducted. In the first version, the dot probe appeared immediately after the cues (as is more common in spatial cueing tasks: MacLeod et al., 1986), rather than after a jittered delay. In the second one, along with having no delay between cue and probe, the cues were also only displayed for 100ms (Cooper & Langton, 2006).

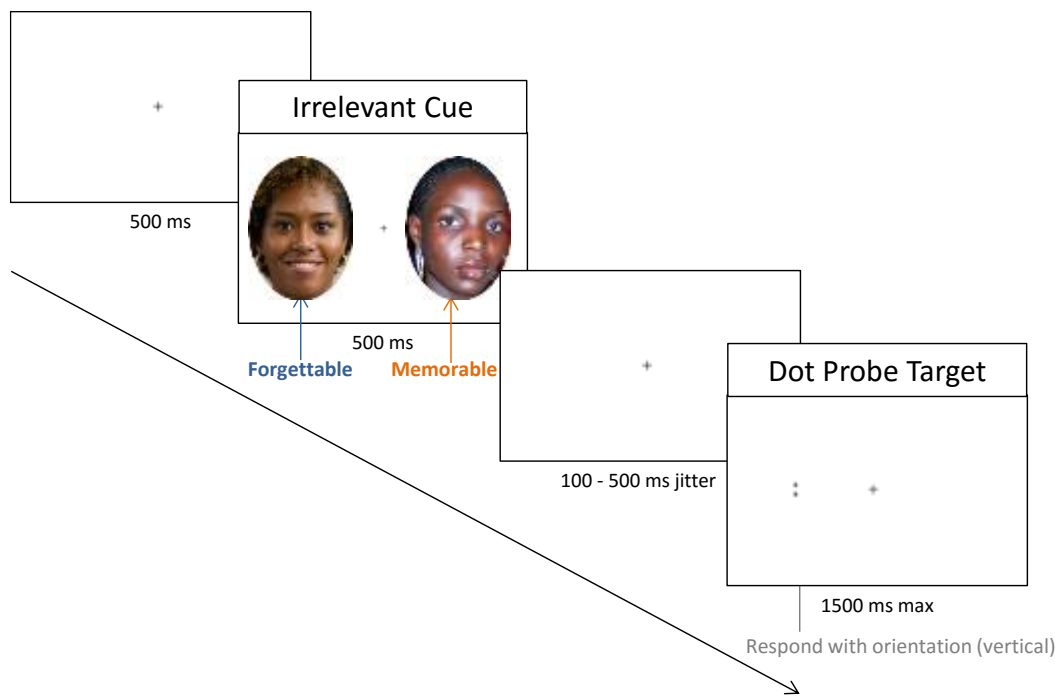

**Fig. 1.** The spatial cueing experimental paradigm for Experiment 1. Participants ignored cues (1-A: one that was very faded and one that was of normal brightness, or 1-B: one that was highly forgettable and one that was highly memorable) and responded with the orientation of a dot probe target. If their response time was influenced more by one type of cue over another, then memorability is likely to automatically capture attention. The face images used in this figure and all other figures in this paper are for illustrative purposes, and are within the public domain.

## Results and Discussion

For Experiment 1-A, reaction times for a target aligned with a visually salient cue (normal brightness, M = 603.17ms, SD = 152.34) were significantly lower than a target aligned with a less salient cue (very bright stimulus, M = 616.83, SD = 168.19), paired t-test: $t(95) = 2.24$, $p = 0.028$. In contrast, no significant difference was found in Experiment 1-B for reaction times for targets aligned with memorable cues (M = 619.13ms, SD = 107.21) or forgettable cues (M = 624.62, SD = 115.09), $t(97) = 1.03$, $p = 0.306$. As a non-significant p-value does not indicate support for the null hypothesis, two-tailed Bayesian hypothesis testing was used to examine if these data support the null hypothesis (Gallistel, 2009; Jarosz & Wiley, 2014), where memorable cues and forgettable cues would indeed have no difference. This analysis uses Bayesian Factor analysis to determine the ratio of evidence (i.e., "odds") in favor of one prior (the null distribution) versus an alternative prior (the largest plausible effect size; the effect size found in Experiment 1-A). The resulting odds (i.e., a Bayes factor, $BF_{01}$) were 1.11:1 in favor of the null hypothesis, providing mild evidence that memorable and forgettable cues do not result in a different reaction time in a spatial cueing task.

Different timings in Experiment 1-B did not cause an effect to appear; there was still no significant difference between memorable and forgettable image cue RTs when the random delay was removed ($t(25) = 0.45$, $p = 0.660$; $BF_{01} = 1.08$) or when the cue presentation time was changed to 100ms ($t(22) = 0.67$, $p = 0.508$; $BF_{01} = 1.06$).

As expected, Experiment 1-A was able to successfully replicate previous effects found in spatial cueing paradigms based on a low-level visual feature (Johannes et al., 1995; Lupiáñez et al., 2004) that results in attentional capture. In contrast, Experiment 1-B did not find a link between memorability and attentional capture, even when using the same timings used by studies able to find attentional capture by emotional faces (Cooper & Langton, 2006). These results implicate memorability as a phenomenon separate from bottom-up attention; these highly memorable stimuli do not appear to contain low-level visual features that are automatically capturing attention. However, it is possible that another bottom-up attentional task may be better suited for uncovering attentional effects on memorability.

## Experiment 2: Memorability in Visual Search

**Introduction**

There has been debate as to whether spatial cueing attentional effects are robust and replicable (Staugaard, 2009), and its strongest effects with faces are often found in special populations (Asmundson & Stein, 1994; Bar-Haim et al., 2007), or with strong threat-based biases (Trawalter et al., 2008). In contrast, a visual search paradigm may provide a more nuanced understanding of the interplay of memorability and attention, and lend evidence as to whether memorability is an attention-driven stimulus property. Do memorable targets quickly capture attention, and are thus easily identified? Do memorable distractors capture attention and make it harder to zero-in on a target? Previous work has found that it is easier to find a deviant amongst standard stimuli than vice versa (Treisman & Gormican, 1988; Wolfe, 2001). As memorable stimuli have been found to be generally distinctive (Bainbridge et al., 2013), it is possible that memorable stimuli may show the same pattern.

In order to examine these questions, a face image visual search experiment was conducted online, with targets and distractors of varying memorabilities. If memorability captures attention, we should anticipate that memorable targets will be very quick to be identified, but also that memorable distractors will detract attention from the visual search. However, if memorability does not capture attention, then we would not see a meaningful effect of target or distractor type.

## Materials and Methods

*Participants*

Participants were recruited on AMT and compensated for their time. Ultimately 74 Workers participated in the study. Only participants with over a 95% AMT approval rating and an IP address in the United States were recruited for the study.

*Stimuli*

The experiment used 180 highly memorable and 180 highly forgettable face images, from the set described in Experiment 1 Materials and Methods.

*Experimental Methods*

The experiment was coded and conducted using PsyToolkit (see Fig. 2). The stimuli were grouped into 32 conditions that varied along four factors: 1) whether the target was present or absent, 2) whether the target was memorable or forgettable, 3) whether the distractors were memorable or forgettable, 4) search set sizes of 4, 8, 12, or 16 stimuli. Participants were asked to respond as quickly and accurately as possible whether a target was present or absent with a key press for both responses.

For each trial, the target to search for was presented above the search display for 1500ms. Then, a search display as a $4 \times 4$ grid (similar to the visual search display of Golan et al., 2014) appeared below the target, separated by a horizontal line. The number of images in the grid was determined based on the set size of that trial (4, 8, 12, or 16), and were placed in randomized locations (with unused locations blank). On target present trials, the target was placed in a random location in the grid amongst distractors, while on target absent trials, only distractors were used. The target (if present) was either taken from the highly memorable or highly forgettable set, and the distractors were taken from either set, based on condition. The specific images used were selected randomly.

Participants were given 5000ms to make their response of target present or target absent and RT was measured. They were given feedback for 1000ms after every response. A noise mask was displayed for 200ms, and then there was a rest between trials for 2000ms. The target cue appeared before the search grid and remained on for the whole trial to diminish any memory-related effects on performance. That is, even if a forgettable target may not cause lesser attentional capture compared to a memorable target, we could find a slower search response because participants are having a harder time remembering what the target looked like. With the target always available to the participant and with time before each trial to encode the target, ideally participants should be able to retain the target in memory for the duration of the trial. Participants completed 32 trials (one per condition), and the experiment took approximately 3 minutes in total. Only trials where participants responded correctly on the task (target absent / present) were used in the analyses. Analyses were conducted using two methods: 1) looking at RT using repeated measures fixed effects linear mixed models (target presence, target memorability, and distractor memorability as categorical factors; set size as a continuous factor), and 2) looking at visual search slope (e.g., Wolfe, 1998; the slope of a regression line fit to each participants' plots of set size by RT) in repeated-measures ANOVAs.

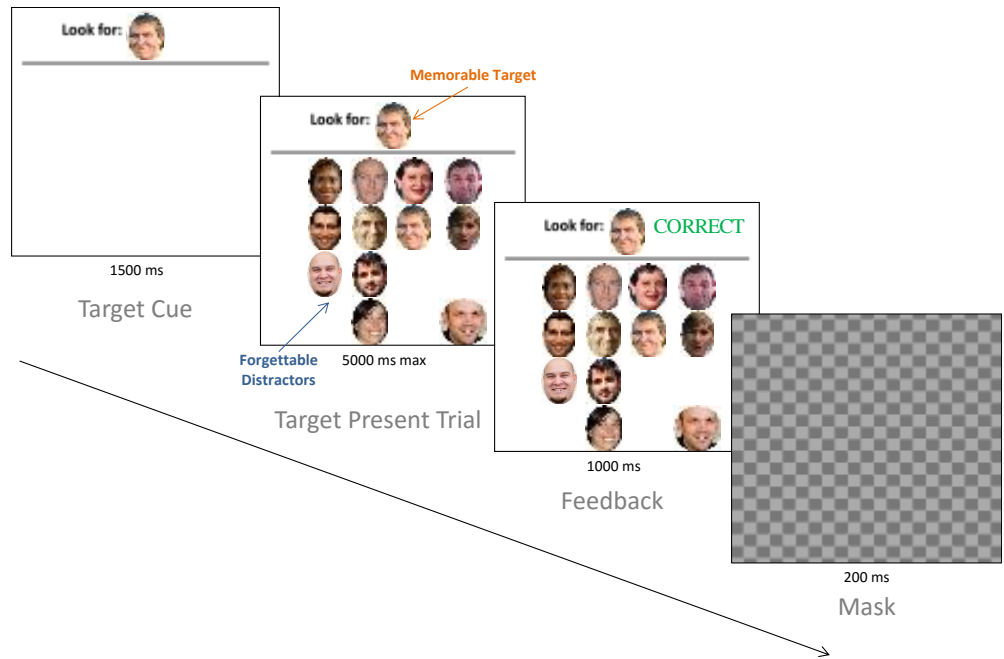

**Fig. 2.** The visual search experimental paradigm for Experiment 2. Participants searched for memorable or forgettable target face images amongst memorable or forgettable distractor images, with different search sizes. In half of the trials the target was present, while in the other half the target was absent. Participants made a response on every trial.

## Results and Discussion

Average RT based on target and distractor memorability can be seen in Fig. 3. As expected from other visual search studies (Treisman & Gelade, 1980), target absent trials took significantly longer to identify than target present trials (linear mixed model modeling main effects: $F = 184.97$, $p = 2.40 \times 10^{-40}$). Analyses were conducted separately for target present and target absent trials, as the interaction of memorability and attention could differ between these two different trial types (as there is no memorable or forgettable target in the target absent trials).

For the target present trials, there was a significant main effect of set size ($\beta = 61.55$, $SE = 8.88$, $F = 176.39$, $p = 5.70 \times 10^{-37}$), with higher set sizes resulting in a longer RT, as expected. There was also a significant main effect of target memorability (linear mixed model: $\beta = 326.19$, $SE = 137.22$, $F = 13.42$, $p = 2.63 \times 10^{-4}$; main effect of target memorability in 2-way ANOVA on slope: $F(1, 68) = 6.36$, $p = 0.014$), with memorable targets generally identified faster than forgettable targets (the red solid lines in Fig. 3A). However, there was no significant main effect of distractor memorability (linear mixed model: $\beta = -186.35$, $SE = 135.78$, $F = 2.61$, $p = 0.107$; 2-way ANOVA on slope: $F(1, 68) = 0.920$, $p = 0.341$; the red solid lines in Fig. 3B). A Bayesian factor analysis also supports the null hypothesis, $BF_{01} = 4.17$, with an alternative prior derived from the RT difference between the target present and absent trials. In terms of statistical interactions, there was no significant statistical interaction of distractor type and target type (linear mixed model: $\beta = 59.26$, $SE = 194.28$, $F = 0.09$, $p = 0.760$; 2-way ANOVA on slope: $F(1, 68) = 0.522$, $p = 0.473$), nor distractor type and set size ($\beta = 23.66$, $SE = 12.76$, $F = 3.09$, $p = 0.079$). However, there was a significant statistical interaction of target memorability and set size ($\beta = -18.01$, $SE = 12.84$, $F = 7.93$, $p = 0.005$). There was no significant 3-way interaction of distractor type, target type, and set size ($\beta = -13.33$, $SE = 18.23$, $F = 0.71$, $p = 0.400$).

a)

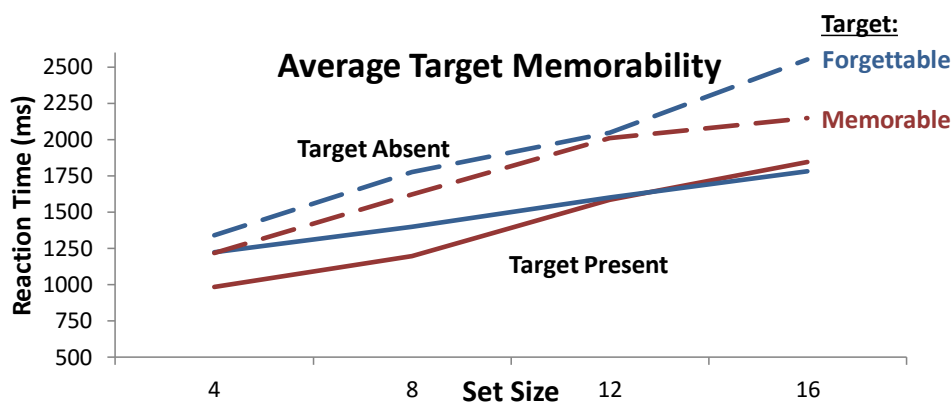

b)

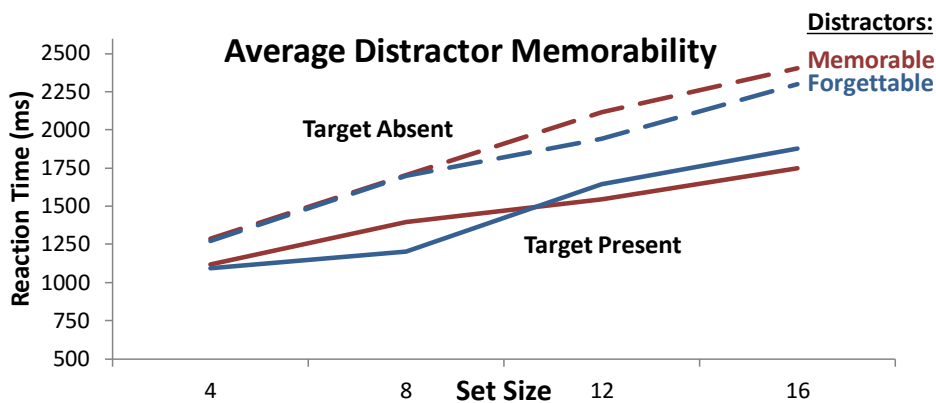

**Fig. 3.** a) The mean reaction times of the conditions averaged by *target memorability*, at each search size. Dashed lines indicate target absent trials, while solid lines indicate target present trials. Red lines indicate memorable target trials, while blue lines indicate forgettable target trials. As expected, target absent trials take longer to identify than target present trials (i.e., dashed lines have longer RTs than solid lines). Larger set sizes also result in longer search times. However, there is no obvious effect of target memorability on search times; while target memorability shows an effect for some set sizes, these effects existed even when the target is absent from the search display.

b) The mean reaction times of the conditions averaged by *distractor memorability*, at each search size. Dashed lines indicate target absent trials, while solid lines indicate target present trials.

Red lines indicate memorable distractor trials, while blue lines indicate forgettable distractor trials. Again, target absent trials as well as larger search set sizes result in longer search times. However, there is no effect of distractor memorability on search times.

To understand the statistical interaction of target memorability and set size for the target present trials, paired t-tests of target memorability were conducted for each set size. Memorable targets only showed a RT advantage for the smaller set sizes of 4 ($t(71) = 3.11$, $p = 0.003$) and 8 ($t(71) = 2.61$, $p = 0.011$), but there was no significant effect of target type for the larger set sizes of 12 ($t(63) = 1.01$, $p = 0.316$) or 16 ($t(63) = -0.57$, $p = 0.569$). (Note that the degrees of freedom differ due to 8 participants who did not successfully respond for any larger set size trials.) These results indicate that memorable targets have an effect on search times only at smaller set sizes.

The target absent trials show a similar pattern (Fig. 3). Based on a three-way linear mixed model (target memorability × distractor memorability × set size) for target absent trials there is again, as expected, a significant main effect of set size ($\beta = 78.73$, $SE = 8.24$, $F = 446.52$, $p = 5.79 \times 10^{-82}$), where it takes participants longer to confirm a target absent trial with more stimuli. There is no significant main effect of target memorability (linear mixed model: $\beta = -143.52$, $SE = 127.13$, $F = 0.21$, $p = 0.650$; 2-way ANOVA on slope: $F(1, 68) = 0.40$, $p = 0.530$), and no significant main effect of distractor memorability (linear mixed model: $\beta = -152.42$, $SE = 126.57$, $F = 0.13$, $p = 0.722$, $BF_{01} = 4.07$; 2-way ANOVA on slope: $F(1, 68) = 2.33$, $p = 0.131$). There is no significant statistical interaction of target memorability and set size ($\beta = 27.40$, $SE = 11.71$, $F = 3.27$, $p = 0.071$) and no significant statistical interaction of distractor memorability and set size ($\beta = 2.77$, $SE = 11.61$, $F = 1.35$, $p = 0.246$). However, there is a trending significant statistical interaction of target memorability and distractor memorability (linear mixed model: $\beta = 369.12$, $SE = 180.87$, $F = 4.17$, $p = 0.042$; 2-way ANOVA on slope: $F(1, 68) = 3.30$, $p = 0.074$), where paired t-tests show that there is a significant target memorability effect when amongst memorable distractors ($t(68) = 2.06$, $p = 0.043$), but not when among forgettable distractors ($t(69) = 0.96$, $p = 0.340$). However, note that this means target memorability is modulating search times, where there is in fact no target present in the search display (i.e., on these target absent trials). There is no significant 3-way statistical interaction of target memorability, distractor memorability, and set size ($\beta = -24.79$, $SE = 16.59$, $F = 2.23$, $p = 0.135$). Looking at paired t-tests of target memorability at each set size, there is an effect of target memorability at the set sizes of 4 ($t(67) = 2.91$, $p = 0.005$), 8 ($t(67) = 2.34$, $p = 0.022$), 16 ($t(68) = 4.35$, $p = 4.75 \times 10^{-5}$), but not at 12 ($t(67) = 0.15$, $p = 0.881$).

While the target present and target absent trials appear to show the same patterns (e.g., target memorability effects at various set sizes, with no significant effect of distractor memorability), these results do not necessarily support a role of memorability in visual search. That is, there is an effect of target memorability (modulated by set size, and in the case of target absent trials, modulating distractor memorability effects), even when *there is no target* in the search display. There is no purely bottom-up attentional explanation independent from memory that could explain how the memorability of a non-present stimulus would influence visual search. The memorability advantage to an absent target is likely instead related to the target search cue; perhaps a memorable cue is easier to hold in memory, so participants do not need to reference back to the cue as frequently during the search. This target memorability advantage is also unstable across set sizes, and may only work for smaller stimulus sets (i.e., where there are fewer images to hold in memory).

Additionally, the lack of effect on search time from distractors of differing memorability provides compelling evidence against memorability being explained by attentional capture. If memorable images indeed capture attention, there should be some detrimental effect on reaction time of memorable distractors flooding the search display, however this is not found in the current study.

<div align="center">Section Discussion</div>

Overall, Experiments 1 and 2 provide strong evidence using two very different paradigms to show that memorability is unlikely to be an image property that causes bottom-up attentional capture. Highly memorable stimuli do not cause automatic spatial cueing, nor do they meaningfully influence visual search times. Memorability as a property cannot be explained by exogenous attentional accounts, and images are not necessarily memorable because they are attention-grabbing.

<div align="center">

## Section 2: Memorability and Top-Down Attention

</div>

<div align="center">Section Introduction</div>

While bottom-up attention is unable to explain memorability, memorability effects could instead occur due to intentional reallocation of attentional resources to stimuli that are later remembered. If memorability is determined by top-down attention, then can participants make themselves remember a forgettable image, or forget a memorable image? Does allocating more attentional resources to a stimulus cause one to remember it better? Previous work has found that cognitive control over memory as well as deeper encoding of certain stimuli over others may work through elaborately encoding distinctive features of a stimulus that one must remember (Eysenck, 1979; Lockhart & Craik, 1990; Basden et al., 1993). Given that memorable images tend to be more distinctive (Bainbridge et al., 2013), perhaps cognitive control of attention results in more elaborate encoding, and thus a better memory for memorable images. Additionally, even if these effects do not fully encompass those of memorability, another important question is to what degree attention may mitigate the effects of stimulus memorability on later memory performance and vice versa. These questions were explored through two classical top-down attentional tasks – directed forgetting (Experiment 3) and a manipulation of encoding depth (Experiment 4) – which ultimately find that while directed forgetting and encoding depth still influence memory performance, memorability causes a stronger, separate effect on memory.

## Experiment 3: Memorability and Cognitive Control

**Introduction**

Previous work has found that participants are able to inhibit explicit memory of a stimulus when given a "directed forgetting" task, where participants are presented with an image or word and then immediately presented with instructions to either remember or forget that stimulus, but then later tested on their recognition memory of all presented images (MacLeod, 1989). It is believed that the memory performance differences between images cued to be forgotten versus remembered is a greater degree of attention and more rehearsal of those cued to be remembered (MacLeod, 2012). Perhaps this greater rehearsal is the key to why some stimuli are consistently more memorable than others.

To answer this question, a directed forgetting task was conducted with stimuli of differing memorability; participants were asked to remember or forget stimuli that were preselected to be of low, medium, or high memorability (unbeknownst to the participant), and then they were tested on their true memory. Depending on the interaction of cognitive control and memorability, there are two possible hypotheses we would expect. First, it is possible that memorability effects are largely explained by top-down cognitive control (i.e., a person decides a memorable image is interesting and decides to encode it). Similarly, it is possible that cognitive control would have a stronger influence on memory than memorability does, as cognitive control has a strong effect on explicit memory (MacLeod, 1989). If either of these are the case, then we should see that cognitive control is the main determinant of ultimate memory behavior, not the memorability of the original image. An alternate hypothesis is that memorability is an intrinsic image property that is unaffected by cognitive control; while people will tend to forget images they try to forget and remember those they try to remember, memorability will have a stronger and separate effect on what they eventually remember and forget.

**Materials and Methods**

*Participants*

Seventy-two participants were recruited on AMT, and screened for having at least a 95% approval rating and an IP address within the United States. Participants were compensated for their time, and given an extra bonus for their performance on the memory test.

*Stimuli*

A set of highly controlled face stimuli of low (bottom 25%), medium (middle 25%), and high memorability (top 25%) were used as stimuli in this study (from the same stimulus set as Experiment 1 Materials and Methods).

*Experimental Methods*

The experiment followed the general methodology of other directed forgetting studies (MacLeod, 1989), and was coded using PsyToolkit. There were two phases to the experiment: a study phase and a test phase (see Fig. 4 for the experimental design). During the study phase, there were 20 stimuli each in a total of 6 conditions, varying along two factors: 1) memorability (low, medium, high), and 2) instructions to the participant (remember / forget), resulting in 120 target stimuli total. For the test phase, there were an additional 120 faces of medium memorability to act as foil faces, with matched statistics with the target faces. Each participant only saw half of the targets and foils (60 images each) to reduce the length of the experiment, so each stimulus was seen by 36 participants.

In the study phase, participants were told that they were going to see a stream of face images, and after each image they would get a cue to either "remember" or "forget" the face. Note that it is important that the cue is after the presentation of the image, so that participants have the same degree of visual experience with each image (in the most extreme case, if the cue comes before the image, they could close their eyes for any image they are told not to encode into memory). Participants were told they would be tested later on their memory and they would get bonus money based on their memory performance. These ambiguous instructions incentivized them to correctly follow the memory cues, as they were unaware that they would ultimately be tested on their recognition for all images. During the study phase, participants saw 60 face images, each one presented for 1000ms, followed by a 2000ms remember or forget cue, and then a 500ms fixation cross. In total, the study phase took approximately 4 minutes.

For the test phase, participants were then tested for their memory. They were told to try and recall everything that they saw (counter to their original expectations), and respond based on whether they had seen the image before or not, regardless of whether they were originally asked to remember or forget it. They were given up to 1500ms to respond to each face which was then followed by a 500ms fixation cross, and they were rewarded with bonus money based on correct responses.

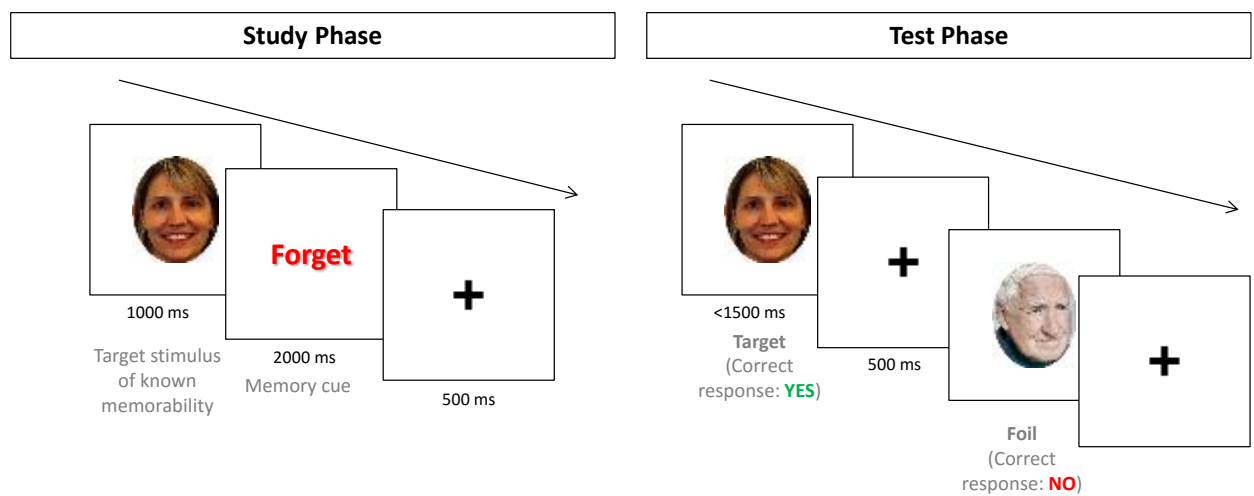

**Fig. 4**. The experimental methods of the study and test phases of the directed forgetting paradigm used in Experiment 3. In the study phase, participants saw a stream of face images (of low, medium, or high memorability) and for each one, were directed to either remember or forget that image, with the incentive of a monetary bonus. In the test phase, participants were told to instead try and remember all of the images they saw in the study phase, regardless of memory cue.

## Results and Discussion

A summary of the main results can be seen in Fig. 5. A 2-way repeated measures ANOVA on participant memory performance during the test phase for the different conditions found a significant main effect emerged of memorability level, $F(2, 426) = 33.93$, $p = 9.02 \times 10^{-}$ [13], with an effect size of $\eta_p^2 = 0.32$. There was also a significant effect of the memory cue, with a lower HR for images participants were told to forget than those they were told to remember ($F(1, 426) = 5.76$, $p = 0.019$), although a smaller effect size of $\eta_p^2 = 0.08$. However, there was no significant statistical interaction between the two factors ($F(2, 426) = 0.26$, $p = 0.775$, $BF_{01} = 7.19$), indicating that directed forgetting does not appear to influence memorability effects.

Looking at specific effects within memorability using paired t-tests, highly memorable images were remembered significantly more than moderately memorable ($t(71) = 4.26$, $p = 6.13 \times 10^{-5}$), and moderately memorable images were remembered significantly more than low memorable ones ($t(71) = 4.59$, $p = 1.88 \times 10^{-5}$).

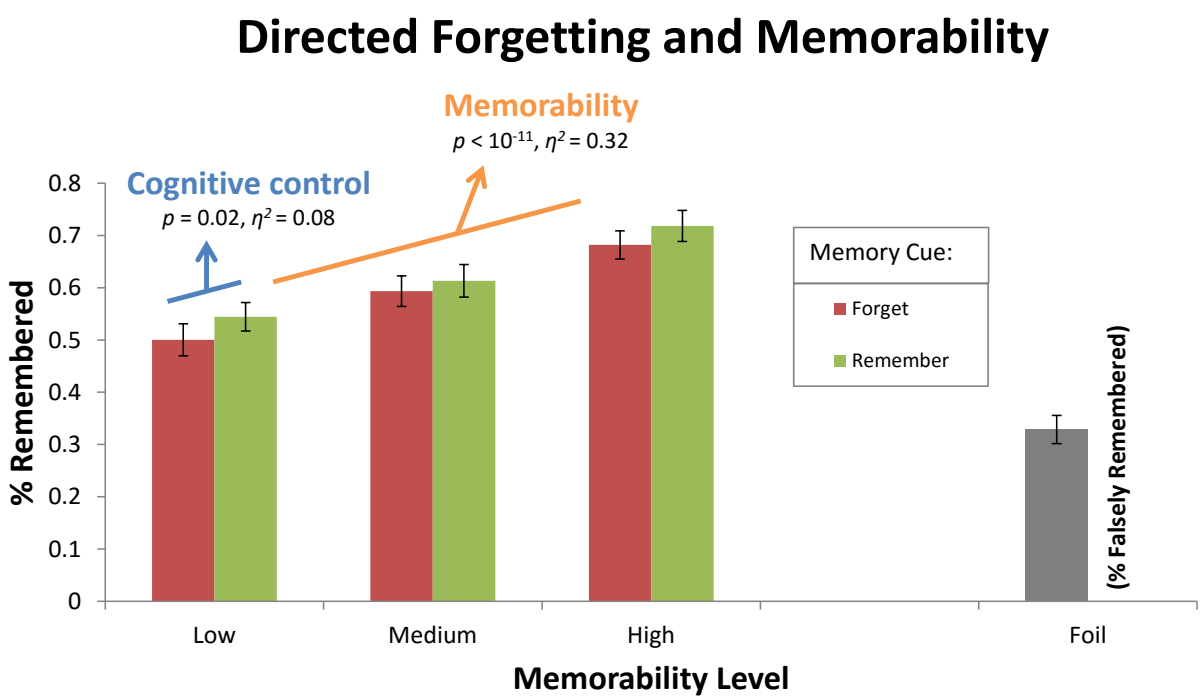

**Fig. 5.** The hit rates of the different conditions, varying along memorability level (low, medium, or high) and memory cue (forget or remember). The false alarm rate for the foil images (all of medium memorability) is also presented as a point of comparison. Error bars indicate standard error of the mean. While there was an effect of the memory cue (people better remembered images they were told to remember than those they were told to forget), image memorability had a significant effect on subsequent memory of larger effect size, with no statistical interaction with directed forgetting.

In sum, these results indicate that people still significantly remembered memorable images over forgettable images, regardless of the memory cue they were presented with at the study phase. At the same time, participants were still correctly performing the task, and the study was able to replicate directed forgetting effects, though with a weaker effect than the effect of memorability. This provides strong evidence that memorability is a relatively immutable property of an image or entity, and that memorability effects cannot be explained by a cognitive control account. Interestingly, just as directed forgetting does not affect implicit memory measures like priming (Vuilleumier et al., 2005), directed forgetting does not alter the influence of memorability on memory performance, providing evidence that memorability could fall along the border of implicit and explicit memory (see Section 3). Essentially, no matter how hard one tries, you cannot make yourself remember a forgettable image, or make yourself forget a memorable image.

## Experiment 4: Memorability and Depth of Encoding

**Introduction**

Another top-down attentional phenomenon that could interact with memorability is depth of encoding, or different levels of processing (Lockhart & Craik, 1990). When stimuli are processed in terms of their semantics or meaning (i.e., deep encoding), they tend to be remembered better than when they are processed in terms of their perceptual features (i.e., shallow encoding) (Bower & Karlin, 1974; Sporer, 1991; Innocenti et al., 2010). This is thought to be due to the greater amount of attentional load and effort required to do deeper processes (Lockhart & Craik, 1990). Memorability effects could thus occur due to deeper encoding or more attentional resources put into remembering memorable images. Perhaps people perform more semantic processing with memorable images (e.g., perhaps they are more interesting or have more semantic content), and thus encode the images deeper.

This question was addressed using an encoding depth task (Bower & Karlin, 1974), where participants categorized sets of memorable and forgettable face stimuli using tasks of three different encoding depths – identifying the color of a fixation cross (shallowest task), the gender of a face (shallow task), or judging the honesty of the face (deep task). Participants were then given an unexpected memory test. If memorability effects occur due to deeper encoding, then controlling for depth of encoding should eliminate a difference between memorable and forgettable images. Alternately, if memorability is intrinsic to images and distinct from encoding depth, we expect to find separate effects of stimulus memorability and task encoding depth on subsequent memory.

## Materials and Methods

*Participants*

Seventy-two participants were recruited on AMT, and screened for having at least a 95% approval rating and an IP address within the United States, to reduce the likelihood of demographics-based biases like the other-race effect (Chiroro & Valentine, 1995). Participants were compensated for their time.

*Stimuli*

A set of highly controlled face stimuli of low and high memorability were used as stimuli in this study (see Experiment 1 Materials and Methods). Faces were used as several encoding depth studies have established paradigms using faces (Bower & Karlin, 1974; Sporer, 1991).

*Experimental Methods*

The experiment followed the general methodology of other depth of encoding experiments (Bower & Karlin, 1974; see Fig. 6). The experiment had four parts, and participants did not know of the different parts beyond the fact that they were related to faces. The first three parts comprised the study phase, using tasks of three different encoding depths where participants had to make different binary decisions on the face images, and the fourth part was an unexpected test phase. For the shallowest processing task, participants were asked to identify the color (black or white) of a fixation cross that appeared on the face image (the "fixation task"). For a deeper task, participants were asked to identify the gender (male or female) of a face image (the "gender task"). This task is often used as the shallow processing task in depth of encoding experiments (Bower & Karlin, 1974), however as gender determination requires holistic face processing, it is likely that it is "deeper" than the fixation cross task. Lastly, for the deepest task, participants were asked to judge how honest (honest or dishonest) they thought a face was (the "honesty task"), as used in other work (Bower & Karlin, 1974). All tasks had a black or white fixation cross on each face (with color distributed evenly over memorable and forgettable images), so stimuli were visually identical and differed only in task and randomized image set. Forty target face stimuli were used in each task, with half being highly memorable images and the other half highly forgettable, resulting in 120 target stimuli total. However, each participant only saw half of the stimuli (60 images) to reduce the length of the experiment, so each stimulus was seen by 36 participants. Each image was displayed for 1000ms and was separated by a 500ms fixation cross, for a total time of 30s per part. The order of these three tasks was counterbalanced across participants, images were randomly sorted into each task, and participants were asked to focus only on the task at hand and not think about the other tasks they had completed.

The fourth part for all participants consisted of an unexpected memory test phase. Participants saw a stream of images and were told to identify which they had seen earlier in the experiment. Sixty of the images were targets, while 60 were foils, and they were presented in a randomized order. Participants were given up to 1500ms to respond to each face which was then followed by a 500ms fixation cross. Both reaction time and performance were recorded. The experiment took approximately 5 minutes in total.

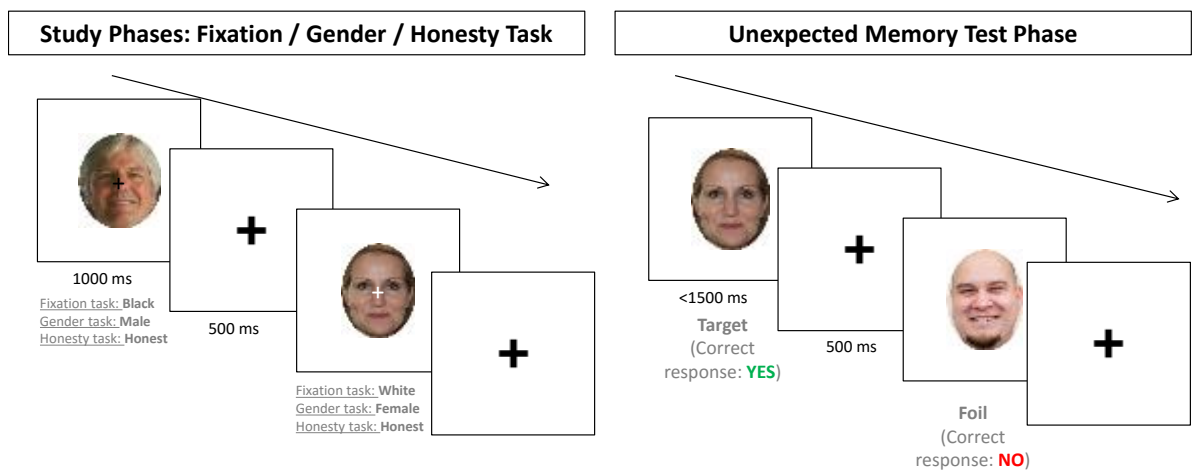

**Fig. 6**. The depth of encoding experimental design of Experiment 4. The experiment consisted of four parts. The first three were identical in paradigm, but had three different tasks, counterbalanced in order across participants: 1) the fixation task, 2) the gender task, and 3) the honesty task. Displayed are example responses that would be indicated based on the task. After the three study parts, participants then completed an unexpected memory test on all the stimuli that were presented in the three previous parts.

**Results and Discussion**

A graphical summary of the results can be seen in Fig. 7. Based on a 2-way within-subjects repeated-measures ANOVA (memorability × encoding depth) on RTs, participants responded significantly faster to memorable images in the memory test than to forgettable images ($F(1, 426) = 10.87$, $p = 0.002$), with an effect size of $\eta_p^2 = 0.13$. There is also a marginally significant main effect of the encoding task on RT ($F(2, 426) = 2.98$, $p = 0.054$, $\eta_p^2 = 0.04$), with faster reaction times for images that were studied with deeper encoding. However, there is no significant statistical interaction between memorability and encoding depth with RT ($F(2, 426) = 0.94$, $p = 0.395$, $BF_{01} = 5.30$).

Based on a paired t-test, RTs during the memory test to memorable images were significantly different from those to foil images ($t(71) = 3.30$, $p = 0.002$), however forgettable image RTs were not different from those of foils ($t(71) = 0.27$, $p = 0.790$, $BF_{01} = 1.09$). When looked at by task using paired t-tests, RTs in the memory test were not significantly different between memorable and forgettable images for the fixation task ($t(71) = 1.02$, $p = 0.313$), however they were for the gender task ($t(71) = 2.06$, $p = 0.043$) and the honesty task ($t(71) = 3.24$, $p = 002$).

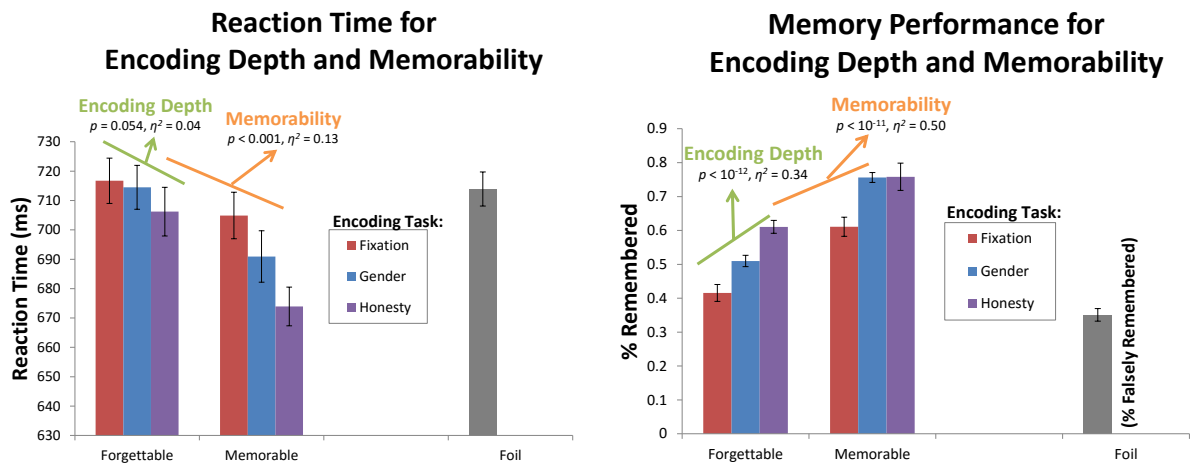

**Fig. 7**. (Left) Reaction time on the unexpected recognition memory test based on the different conditions. Memorable images had significantly faster reaction times than forgettable images. There is also a significant effect of task encoding depth, with deeper tasks causing faster recognition, however there is no statistical interaction between memorability and encoding depth. The reaction time to respond to foil images is comparable to that of forgettable images. (Right) Hit rate on the unexpected recognition memory test based on the different conditions. Memorable images had significantly higher hit rates than forgettable images. Similarly, greater encoding depth also resulted in higher hit rates, however there was no statistical interaction between memorability and encoding depth. The bar for the foil images here reflects false alarm rate, for a point of comparison. Error bars indicate standard error of the mean.

Participant performance in the memory test mirror these RT results. In a 2-way repeated measures ANOVA of memorability and encoding depth, there is a significant main effect of memorability on HR ($F(1, 426) = 70.73$, $p = 2.91 \times 10^{-12}$, $\eta_p^2 = 0.50$). There is also a significant main effect of task encoding depth on HR ($F(2, 426) = 36.32$, $p = 1.83 \times 10^{-13}$, $\eta^2 = 0.34$), with smaller effect size, where images that were encoded with a deeper task show a higher HR. There was also a significant statistical interaction between memorability and encoding depth ($F(2,426) = 4.77$, $p = 0.01$).

Paired t-tests were used to investigate specific differences between the conditions and this interaction effect. Memorable images were remembered significantly more often than forgettable images on all of the tasks (paired t-tests; fixation task: $t(71) = 6.37$, $p = 1.61 \times 10^{-8}$; gender task: $t(71) = 8.12$, $p = 1.02 \times 10^{-11}$; honesty task: $t(71) = 5.23$, $p = 1.61 \times 10^{-6}$). Looking at paired t-tests based on the encoding task, for forgettable images, performance was significantly higher for the gender task than the fixation task ($t(71) = 3.98$, $p = 1.62 \times 10^{-4}$), and higher for the honesty task than the gender task ($t(71) = 3.84$, $p = 2.67 \times 10^{-4}$). For memorable images, performance was significantly higher for the gender task than the fixation task ($t(71) = 5.63$, $p = 3.33 \times 10^{-7}$), but there was no difference for the honesty task compared to the gender task ($t(71) = 0.10$, $p = 0.923$). This is likely due to the fact that performance for these two tasks for memorable images is essentially at ceiling; when told to explicitly remember these images (see Experiment 3; these are the same image sets), participants have about the same performance (gender task $M = 0.76$, honesty task $M = 0.76$, explicit memory $M = 0.73$). This ceiling effect also likely explains the statistical interaction between memorability and encoding depth.

These results show strong effects of both memorability and encoding depth on subsequent memory. In fact, performance was significantly better for memorable than forgettable images on all tasks, and memorability effects had higher effect sizes than encoding depth effects. This indicates that controlling for encoding depth does not equalize memorability; even if you are encoding a set of images deeply and semantically, you will still remember memorable images more than forgettable images. Or, similarly, even when focusing on an irrelevant perceptual item (i.e., fixation crosses overlaid on the images), you will still remember memorable images more than forgettable images. These results imply that effort, distribution of attentional resources, or elaboration of encoding are unlikely to explain the phenomena we find with memorability.

<u>Section Discussion</u>

Overall, Experiments 3 and 4 show that while intentional memory encoding and deeper processing do indeed boost memory performance, these boosts are separate from and often weaker than the impact of the intrinsic memorability of the stimulus to be remembered. Memorability is thus neither equivalent to bottom-up attention, nor top-down attention. As it seems that memorability is not an attention-based phenomenon, these results also provide evidence that memorability may not be an explicit memory phenomenon, as cognitive control is known to be able to influence explicit memory, but not implicit memory (MacLeod, 1989; Vuilleumier et al., 2005). One final important question, then, is how memorability might relate to an implicit memory process in the absence of attention.

# Section 3: Memorability and Priming

## Introduction

As memorability appears to be a nonconscious, implicit, automatic form of memory that is independent of attentional effects, how is it linked to perceptual priming, a similarly automatic and nonconscious form of memory? Like memorability, perceptual priming has been shown to be unaffected by changes in low-level visual features (Fiser & Biederman, 2001) and top-down attention (Vulleumier et al., 2005). Since memorability seems to happen at the timing of perception but does not capture bottom-up attention, memorability might thus reflect the "primability" of a stimulus – to what degree behavioral and neural responses are affected by increasing repetitions of an initially novel stimulus. Memorable images might be those that cause greater priming effects, while forgettable images show less priming.

To test the link between memorability and "primability", a perceptual priming experiment was conducted, where participants had to rapidly categorize scene images for indoor / outdoor (Experiment 5-A) or natural / manmade (Experiment 5-B). Images were repeated four times each, but with the repetitions spread across the stimulus presentation stream in a randomized order. Due to perceptual priming, with increasing repetitions a stimulus will become easier (and faster) to categorize. If memorability and primability are linked, then memorable images should show a more pronounced drop in reaction time with each repetition, in comparison to forgettable images. However, if memorability and primability are separate phenomena, then the memorability of the stimulus should not affect priming effects.

## Materials and Methods

### Participants

Forty-nine participants recruited from AMT participated in Experiment 5-A, and a separate set of 48 participants participated in Experiment 5-B. They were selected for having at least a 95% approval rating, and were compensated for their time.

### Stimuli

Scene images from a highly controlled stimulus set for both low-level visual features (e.g., color, spatial frequency) and higher-level attributes (e.g., number of objects, average object size) were used in the two experiments (Xiao et al., 2010; Isola et al., 2013). Scene images were used here as they can be easily and quickly categorized for multiple category dichotomies (e.g., indoor / outdoor, natural / manmade), yet do not have the potential demographic-based categorization biases that could occur with faces, such as the other-race effect (Chiroro & Valentine, 1995) or other-age effect (Anastasi & Rhodes, 2005). All images were color and 256 pixels × 256 pixels.

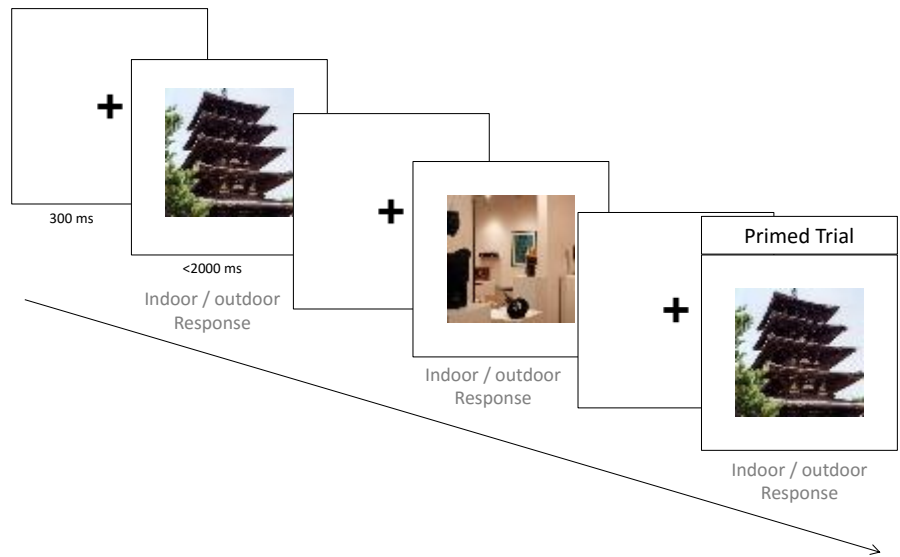

**Fig. 8**. The perceptual priming experimental paradigm for Experiment 5. Half of the images were highly forgettable, while the other half were highly memorable. Participants responded as quickly as possible to a perceptual categorization task (indoor / outdoor for Experiment 5-A, natural / manmade for Experiment 5-B) for a stream of images, where sometimes an image would repeat. On these repetition trials, we can observe the effects of perceptual priming (reaction time decreasing on repeated stimuli), and if this differs between memorable and forgettable stimuli.

For Experiment 5-A, the scene images varied along two factors with 4 conditions total, with 12 stimuli each, or 48 stimuli total: 1) memorable or forgettable, and 2) indoor or outdoor. Experiment 5-B had the same stimulus condition distributions, except its second factor was natural or manmade, and all images were outdoor scenes. The images for the two stimulus sets were selected from the same larger scene stimulus set.

*Experimental Methods*

Both experiments were programmed and conducted using PsyToolkit and followed the same experimental paradigm (Fig. 8). For each trial, a fixation cross was displayed for 300ms. A scene image was then presented at central fixation, and participants were given 2000ms to classify the image as indoor or outdoor in Experiment 5-A or natural or manmade in Experiment 5-B with a key press, with reaction time recorded. Each image was repeated four times over the course of the experiment in a randomized order, although participants were not told in advance that they would see image repetitions. Participants were informed if they responded incorrectly, or took too long (over 2000ms) to respond, to encourage quick and accurate responses. For both experiments, participants completed 192 randomized order trials, which took approximately 3 minutes in total. Only trials with the correct task responses were used in the analyses.

**Results and Discussion**

*Experiment 5-A: Indoor / Outdoor Task*

A graphical summary of the results can be seen in Fig. 9. A 2-way repeated-measures ANOVA (memorability × repetition number) was conducted on the reaction times. As expected based on previous perceptual priming work (Wiggs & Martin, 1998; Turk-Browne et al., 2006), with increasing repetitions of an image, participants were able to more quickly identify it as indoor / outdoor ($F(3, 184) = 29.23$, $p = 2.60 \times 10^{-12}$). However, memorable and forgettable images had no significant difference in how long it took to classify them as indoor / outdoor ($F(1, 184) = 0.002$, $p = 0.968$, $BF_{01} = 4.89$). There was also no significant statistical interaction between the two factors for RT ($F(3, 184) = 1.54$, $p = 0.213$, $BF_{01} = 2.26$), indicating that forgettable and memorable images did not appear to experience different degrees of priming. A linear mixed model modeling memorability as a categorical factor and repetition number as a continuous factor shows the same patterns; RT speeds up with more repetitions ($\beta = -19.88$, $SE = 3.09$, $F = 104.65$, $p = 2.13 \times 10^{-19}$), but memorability shows no effect ($\beta = 12.20$, $SE = 11.96$, $F = 1.04$, $p = 0.309$), nor is there a statistical interaction between memorability and repetition number ($\beta = -4.93$, $SE = 4.37$, $F = 1.27$, $p = 0.261$). Based on paired t-tests, forgettable images and memorable images showed no significant RT differences at any repetition number (all $p > 0.05$). Thus, while scene images do show perceptual priming, there appears to be no differences between memorable and forgettable images.

*Experiment 5-B: Natural / Manmade Task*

　　To fully ensure there is no interaction of memorability and primability, the study was replicated using a different categorization task (natural / manmade), see Fig. 9. Again, as expected, there was a significant effect of image repetition on classification speed ($F(3, 184) = 13.39$, $p = 5.55 \times 10^{-7}$). However, there was again no significant effect of memorability ($F(1, 184) = 4.14$, $p = 0.054$), although in the Bayesian Factor analysis, there was more evidence for the alternate hypothesis than the null hypothesis: $BF_{01} = 0.67$, indicating unclear evidence for whether memorability has an effect on natural / manmade responses. However, importantly there was no statistical interaction between image repetition and memorability ($F(3, 184) = 1.27$, $p = 0.293$, $BF_{01} = 2.58$), indicating that the effect of memorability does not change with priming. A linear mixed model modeling memorability as a categorical factor and repetition as a continuous factor finds a significant effect of repetition ($\beta = -11.46$, $SE = 3.89$, $F = 30.56$, $p = 1.22 \times 10^{-7}$), but no effect of memorability ($\beta = 9.87$, $SE = 15.08$, $F = 0.43$, $p = 0.514$) nor a statistical interaction of memorability and repetition ($\beta = -7.53$, $SE = 5.51$, $F = 1.87$, $p = 0.174$). Looking at the difference between memorable and forgettable images at each repetition using paired t-tests, memorable images were significantly faster to classify than forgettable images at the forth repetition ($t(47) = 3.31$, $p = 0.002$), however there were no significant differences at the first, second, or third presentations of the image (all $p > 0.4$). Again, this study shows no strong evidence for a differential priming effect between forgettable and memorable images.

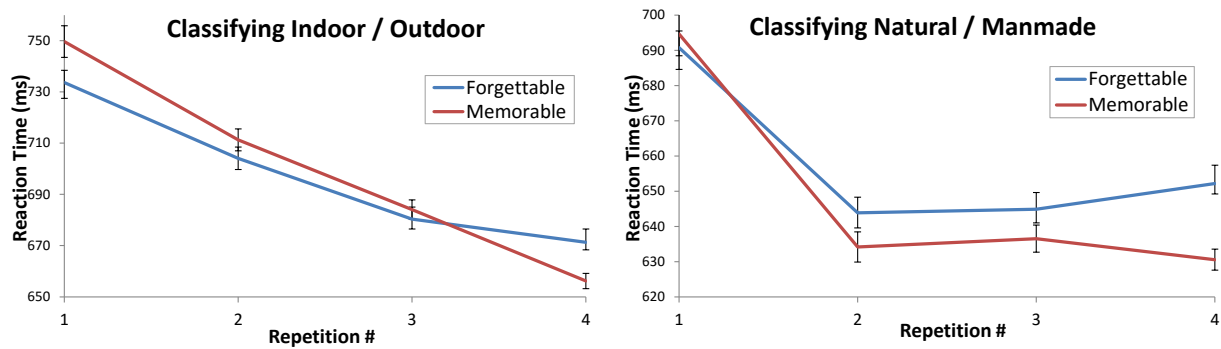

**Fig. 9**. (Left) The mean reaction time for forgettable versus memorable scenes in Experiment 5-A (indoor / outdoor), by repetition number. Perceptual priming occurred equally for forgettable and memorable images. (Right) The mean reaction time for forgettable versus memorable scenes in Experiment 5-B (natural / manmade), by repetition number. Again, there was no significant effect of memorability, nor an interaction of perceptual priming effect and memorability, showing that memorability is unlikely to be equivalent to "primability". Error bars indicate standard error of the mean.

Both Experiments 5-A and 5-B replicate the finding that while people become faster at categorizing scenes (for either indoor / outdoor or natural / manmade) with increasing repetitions of an image, this speed increase (or "primability" of the stimulus) does not appear to be related to memorability. If this were the case, then one would expect to see a significant statistical interaction between repetition time and image memorability. This being said, memorable images were classified significantly faster in the fourth presentation in the natural / manmade categorization task. Because this effect did not exist across the two tasks, it is unlikely that it is due to differential perceptual priming and memorability. Ultimately, both Experiments demonstrate that memorability is its own automatic, intrinsic image property, separate from perceptual priming effects.

These results bring up an interesting question of whether memorability can be considered an implicit or explicit memory phenomenon. These current results show evidence that memorability does not resemble other common implicit memory phenomena, such as priming, despite being a similarly automatic, unconscious marker of memory. However, memorability is also different from explicit memory phenomena, like found in individual explicit recall, because it is unaffected by bottom-up and top-down attention (Vuilleumier et al., 2005). Previous work has found a neural dissociation between implicit memory (i.e., perceptual priming) and explicit memory (i.e., subsequent memory), where implicit memory shows repetition suppression and explicit memory shows pattern similarity (Ward et al., 2013). Neural signatures of memorability have been identified in many memory-related regions, such as the medial temporal lobe (Bainbridge et al., 2017), and future work will need to investigate whether memorability somehow bridges this dissociation of implicit and explicit memory, or could serve as an interesting test case of how we define these two phenomena.

## General Discussion

This set of five experiments shows strong evidence that memorability effects cannot be explained away by attention-related phenomena known to usually interact with memory. Namely, we find that:

Memorability does not cause bottom-up attentional capture, as memorable images do not produce biases in a spatial cueing task (Experiment 1), nor in a visual search task (Experiment 2).

Memorability is not affected by top-down attention, as you cannot make yourself forget a memorable image, or remember a forgettable image (Experiment 3), and memorability effects exist independent of how deeply stimuli are encoded (Experiment 4).

Memorability is not the same as priming – although these are both automatic, unconscious forms of memory independent from attention, these are two separate memory phenomena (Experiment 5).

These results provide further evidence that memorability is an intrinsic, isolated property to an image or an entity (Isola et al., 2011a; Bainbridge et al., 2013), that goes beyond low-level visual features that automatically capture attention. The perceptual features that cause images to be memorable are likely much more subtle and later in the visual processing hierarchy, as evidenced by deep-learning models of memorability (Khosla et al., 2015), and will need further investigation (though with some preliminary findings in Isola et al., 2011b; Bainbridge et al., 2013). Additionally, memorability may be pinpointing a novel subprocess of memory, as it is resilient to attentional effects, unlike explicit memory encoding, which is often considered intertwined with attention (Mulligan, 1998; Kim, 2011). In fact, even explicitly attempting to *not* encode a memory still results in memorability effects on later retrieval. Perhaps the only way to escape the strong effects of memorability may be through direct manipulation of the image, to diminish the features that make it memorable (Khosla et al., 2013). At the same time, these results also show that memorability may not be a purely implicit memory phenomenon, as it is independent from priming effects. The question of memorability as either an implicit or explicit memory phenomenon is an extremely interesting one that may challenge notions of the two types of memory being completely dissociable (Ward et al., 2013). As shown by these five Experiments, memorability is extremely easy to adapt to classical behavioral and neuroscientific paradigms, as it is only a matter of measuring and selecting stimuli, and several memory studies could be reanalyzed with a focus on stimulus memorability. Thus, there is a large range of possibilities for further investigations on the role of memorability in terms of implicit versus explicit memory, bottom-up versus top-down processing, and perception versus memory.

Overall, attention is not a determinant of memorability. While attention-grabbing stimuli may tend to be more memorable, these are two isolable phenomena; images do not need to be attention-grabbing or cause implicit priming to ultimately be more memorable. Memorability is also unwavering, regardless of the effort or depth with which you process images. In whole, memorability has shown itself to be an independent, intrinsic property to images serving as a strong determinant of what we will ultimately remember.

## Acknowledgments

Great thanks to Aude Oliva for her support and advice with this work. W.A. Bainbridge is supported by the Department of Defense, through the National Defense Science & Engineering Graduate (NDSEG) Fellowship Program, and by a teaching fellowship at the Department of Brain and Cognitive Sciences, MIT. The raw data and experimental code for this study will be made publicly available on the author's (Wilma A. Bainbridge) website: http://www.wilmabainbridge.com/